\documentclass[citeauthoryear]{apinv}

\usepackage[T1,T2A]{fontenc}
\usepackage[utf8]{inputenc}
\usepackage{amsmath, amsfonts}
\usepackage{graphicx}
\usepackage{epsfig,cite,graphics}
\usepackage{subfig}
\usepackage{tabularx}
\usepackage{epstopdf}

\begin{document}

\title{Compact statis stars with polytropic equation of state in minimal dilatonic gravity}
\titlerunning{Compact statis stars with polytropic equation of state in minimal dilatonic gravity}
\author{Plamen Fiziev\inst{1,2}, Kalin Marinov\inst{3}}
\authorrunning{P. Fiziev, K. Marinov}
\tocauthor{Plamen Fiziev, Kalin Marinov}
\institute{Sofia University Foundation of Theoretiacal and Computational Physics and Astrophysics, Boulevard 5 James Bourchier, Sofia 1164, Bulgaria
	\and BLFT, JINR, Dubna, 141980 Moskow Region, Russia
	\and Sofia University Foundation of Theoretiacal and Computational Physics and Astrophysics, Boulevard 5 James Bourchier, Sofia 1164, Bulgaria   \newline
	\email{fiziev@phys.uni-sofia.bg},
	\email{fiziev@theor.jinr.ru}	\newline
	\email{marinov.kalin@gmail.com} }
%\papertype{Submitted on xx.xx.xxxx; Accepted on xx.xx.xxxx}	
\maketitle

\begin{abstract}
We present solution of the equations for relativistic static spherically symmetric stars (SSSS) in the model of minimal dilatonic gravity (MDG) using the polytropic equation of state. A polytropic equation of state, which has a good fitting with a more realistic one, is used. Results are obtained for all variables of a single neutron star in the model of MDG. The maximum mass about two solar masses is in accordance with the latest observations of pulsars. Several new effects are observed for the variables related with the dilaton $\Phi$ and the cosmological constant $\Lambda$. The mass-radius relation is also obtained. Special attention is paid to the behavior of the quantities which describe the effects analogous to those of dark energy and dark matter in MDG. The results of the present paper confirm the conclusion that the dilaton $\Phi$ is able to play simultaneously the role of dark energy and dark matter.
\end{abstract}

\section*{Introduction}
The minimal dilatonic gravity (MDG) model is a proper generalization of the Einstein general relativity (GR). Its purpose is to solve some of the problems, which  appear when one tries to describe the Nature using GR as a theory of gravity and the Standard Model of particle physics, and could be a part of a more general theory. It was introduced for the first time by O'Hanlon (O'Hanlon,1972) without relation with cosmology and astrophysics. His point was to give some field-theoretical basis for the Fujii's "fifth force".
An additional field $\Phi$ was introduced there and the term "dilaton" was used for the field $\Phi$.
The idea of a possible relation of the O'Hanlon model with astrophysics and cosmology was introduced and developed in the articles (Fiziev, 2000), (Fiziev, 2002), (Fiziev, 2003), (Fiziev, 2013), (Fiziev, 2014a), (Fiziev, 2014b),where the cosmological constant $\Lambda$ was bring into use in MDG.
This model is based on the following action for the gravi-dilaton sector
\begin{equation}
S_{g,\Phi}=-\frac{c}{2k}\int \mathrm{d}^{4}x \sqrt{\vert g\vert}(\Phi R+2\Lambda U(\Phi)) ,
\label{MDG}
\end{equation}
where  $k=8\pi G/c^{2}$ is the Einstein constant, $G$ is the Newton gravitational constant, and $ \Phi\in (0,\infty) $ is the dilaton field. The values of $\Phi$ must be positive since the change of the sign would lead to the change of the sign of the gravitational factor $G/\Phi$, which leads to antigravity. The value $\Phi =\infty$ must be excluded, because the gravity is turned off. $\Phi =0$ is also unacceptable since it leads to an infinite gravitational factor, and the Cauchy problem is not well posed.
\par
The function $U(\Phi)$ defines the cosmological potential. It must be a positive single value function of the dilaton field by astrophysical reasons. All the physical requirements for the cosmological potential $U(\Phi)$, necessary for a sound MDG model, can be found in (Fiziev, 2013). A class of withholding potentials is introduced there. These confine dynamically the  values of the dilaton $\Phi$ in the physical domain.
\par
The introduced scalar field $\Phi$ leads to a variable gravitational factor $G(\Phi)=G/\Phi$ instead of the gravitational constant $G$. The cosmological potential $U(\Phi)$ is introduced to consider a variable cosmological factor instead of the cosmological constant $\Lambda$. In GR, with the cosmological constant $\Lambda$ we have $\Phi \equiv 1$ and $U(1)\equiv 1$. Due to its specific physical meaning, the field $\Phi$ has unusual properties.
\par
The MDG without a cosmological term corresponds to the Brans-Dicke theory with an identically vanishing parameter $\omega$.
The MDG is only locally equivalent to the f(R) theories and, in general, yields different physical consequences (Fiziev, 2013).
More information about the f(R) theories can be found in the following publications (Clifton, 2012), (De Felice, 2010), (Faraoni, 2006), (Frolov, 2008), (Nojiri S., 2007), (Nojiri S., 2011), (Sotiriou, 2010), (Starobinsky, 1980), (Starobinsky, 2007).
%\cite{Clifton, Defelice, Faraoni, Frolov, Sotiriou, Starobinsky, Starobinsky1}
\par
Some physical and astrophysical consequences of MDG are described in
(Fiziev, 2000),(Fiziev, 2002), (Fiziev, 2003), (Fiziev, 2014a), (Fiziev, 2014b).
In particular, in (Fiziev, 2000)  MDG-modifications of the classical GR effects in the solar system:
the Nordtvedt effect, the time delay of electromagnetic waves (the Shapiro effect), and the perihelion shift
were considered.
In the weak field approximation it was shown that MDG is compatible with all known observational
data if the mass $m_{\Phi}$ of dilaton is large enough.
The strongest estimate $m_{\Phi} > 10^{-3}eV$ can be derived from modern data of Cavendish type
experiments at short distances. In terms of the Compton length $l_{\phi}$ this gives
exponential decrease of the dilaton field at distances longer than $100\mu m$.

At present, the experimental fixing of the mass $m_{\Phi}$, or the Compton length $l_{\phi}$
is the most important open physical problem in both MDG and f(R) theories of gravity.
For now, the only estimate is the one from cosmology: $m_{\Phi}\sim 10^{-6} M_{Plank}$ see (Starobinsky, 2007).
It corresponds to the Compton length $l_{\Phi}\sim 10^6 \times l_{Planck}\approx 10^{-29} m$
which is far below that available for current and future experiments.
This value makes hopeless finding some difference between MDG and GR outside the real bodies
of star scales or of smaller dimension.

The problem with different masses of dilaton $\Phi$ may be solved in the framework of MDG using more complicated
cosmological potentials with several minima (Fiziev, 2013) \footnote{There exist much more sophisticated models
like Chameleon one, with more than one scalar field and corresponding nonlinear interactions
designed to display the hypothetical change of the value of the mass $m_{\Phi}$
depending on the surrounding environment.
We prefer to investigate for the beginning only the simplest MDG model as defined by Eq. \eqref{MDG}.}.
Around each minimum we may have very different masses of the scalar field $\Phi$,
or more precisely, of its small deviation $\delta \Phi=\Phi-\Phi_{min}$.
This means that we may have very different coefficients of the quadratic terms in
the Taylor series expansion of the dilatonic potential (see below)
in the vicinity of the corresponding minimum.
These different masses may be seen at different scales,
and may differ essentially from the one derived from CMB cosmological data, which corresponds to the
huge scale of the visible Universe.
Hence, we need to study effects of MDG at different scales (Fiziev 2014b), (Capozziello, 2014).

Thus, we arrive at the idea of studying stars, and especially neutron stars, in the framework of MDG.
The neutron stars are very interesting objects for our purposes, since in them the
strong gravitational fields are known to exist.
Hence, the effects of MDG on the neutron stars structure may be significant.

However, in addition there exists a specific numerical problem.
It turns out that the available at present computer programs are able to work only for (maybe) unrealistic large
values of the Compton length $l_{\Phi}\sim 10 \div 500 km$.
On the other hand, these scales correspond to the scales
of a typical neutron star, and it is interesting to study the effects of MDG on them for corresponding
small masses $m_{\Phi}$, at least as a preliminary study of this issue.

The first such investigation was carried out in (Fiziev, 2014a,b).
There the most idealized equation of state (EOS) of the neutron matter was used,
namely EOS of ideal Fermi gas at zero temperature.
The idea was to compare the MDG and GR results exploring the simplest textbook example
with a clear physical ground and well known physical approximations.
The present paper, being a proper extension of the MS thesis
of Kalin Marinov (Sofia University, October 2014),
considers a more realistic but also idealized example
of a polytropic equation of EOS following (Damour, 1993).

In the literature, one can find several dozens of "realistic" EOS,
based on different physical models of matter in neutron stars.
Unfortunately, at present we are not able to make a decision
what is the right EOS for neutron stars, from both the observational and theoretical point of view.
The two constructed groups of EOS: the soft and the stiff ones,
give still hardly distinguishable results from the observational point
of view in the framework of each group.
In this situation it seems reasonable to study the effects of the change of the very theory of gravity
using the simplest known EOS with a clear physical ground.

The next step: a study of more realistic models for comparison with Nature
will make sense only after revealing the basic new effects of MDG in stars,
and after overcoming the numerical problems with (maybe) more realistic masses $m_{\Phi}$.
Thus, the present paper is just the next step towards realistic models of neutron stars in the MDG.

\section*{1. Basic equations and boundary conditions for SSSS in MDG}

The field equations of MDG with a matter field can be written in the form

\begin{equation}
\Phi G_{\alpha \beta}-\Lambda U(\Phi)g_{\alpha \beta}-\nabla_{\alpha}\nabla_{\beta}\Phi +g_{\alpha \beta}\Box \Phi=\frac{k}{c^{2}}T_{\alpha \beta} ,
\label{MDG1}
\end{equation}

\begin{equation}
\Box \Phi+\Lambda V_{,\Phi}(\Phi)=\frac{k}{3c^{2}}T .
\label{MDG2}
\end{equation}

Here $ T_{\alpha \beta} $ is the standard energy-momentum tensor of the matter, and $T$ is the trace of the same tensor. The dilatonic potential is introduced through the relation  $V_{,\Phi}(\Phi)=\frac{2}{3}(\Phi U_{,\Phi}-2U)= \frac{2}{3}\Phi^{3}\frac{d}{d\Phi}(\Phi^{-2}U)$, more precisely, it is its first derivative with respect to the variable $\Phi$. By standard notation $G_{\alpha \beta}$ is the Einstein tensor.
\par
The static stars are spherically-symmetrical objects with great precision, so we use a space-time interval in the spherical coordinates (Landau, 1975)

%\cite{Landau}
\begin{equation}
\mathrm{d}s^2=e^{\nu(r)}\mathrm{d}t^2-e^{\lambda(r)}\mathrm{d}r^2-r^2\mathrm{d}\theta^2 -r^2sin^{2}\theta \mathrm{d}\varphi^2 ,
\end{equation}
where r is the radial variable. After some more algebra, the equations describing SSSS in MDG are obtained. For the inner domain $r\in [0,r^{\star}]$, where $r^{\star}$ is the radius of the star the structure is determined by the following system:

\begin{equation}
\frac{\mathrm{d}m}{\mathrm{d}r}=\frac{\beta r^{2}\epsilon_{eff}}{\Phi} ,
\label{MDGM1}
\end{equation}
\begin{equation}
\frac{\mathrm{d}p}{\mathrm{d}r}=-\frac{\alpha(p+\epsilon)}{r\big( \Delta-\frac{1}{2}\alpha\beta r^3p_{\Phi}/ \Phi \big) }\Bigg( \frac{\beta r^3 p_{eff}}{\Phi}+m\Bigg) ,
\label{MDGP1}
\end{equation}
\begin{equation}
\frac{\mathrm{d}\Phi}{\mathrm{d}r}=-\frac{\alpha\beta r^2p_{\Phi}}{\Delta} ,
\label{MDGPhi1}
\end{equation}
\begin{equation}
\frac{\mathrm{d}p_{\Phi}}{\mathrm{d}r}=-\frac{p_{\Phi}}{\Delta r}\Bigg( 3r-7\alpha m-\frac{2}{3}\Lambda r^3+\frac{\alpha\beta r^3 \epsilon_{eff}}{\Phi}\Bigg) -\frac{2\epsilon_{\Phi}}{r} .
\label{MDGPPhi1}
\end{equation}

Here the four unknown functions are $m=m(r), p=p(r), \Phi=\Phi (r)$ and $p_{\Phi}=p_{\Phi}(r)$, respectively, the mass, the pressure, the dilaton, and the dilaton pressure. The following indications are used in the system:

\begin{equation}	\begin{split}
\Delta &=r-\frac{2mG}{c^2}-\frac{\Lambda r^2}{3} ,\\
\epsilon_{eff}&=\epsilon+\epsilon_{\Lambda}+\epsilon_{\Phi}, \quad p_{eff}=p+p_{\Lambda}+p_{\Phi} ,\\
\epsilon_{\Lambda}&=\frac{\Lambda}{2\alpha\beta}\big( U(\Phi)-\Phi\big), \quad p_{\Lambda}=-\frac{\Lambda}{2\alpha\beta}\Bigg( U(\Phi)-\frac{\Phi}{3}\Bigg) ,\\
\epsilon_{\Phi}&=p-\frac{1}{3}\epsilon +\frac{\Lambda}{2\alpha\beta}V\rq{}(\Phi)+\frac{\alpha p_{\Phi}\big( \frac{\beta r^3}{\Phi}p_{eff}+m\big)}{2\big( \Delta-\frac{\alpha\beta r^{3}p_{\Phi}}{2\Phi}\big) } .
	\end{split}	\label{9}	 \end{equation}

In the above equations $\epsilon_{eff}, p_{eff}$  denote the effective values of the energy density and pressure, and they combine in them the matter energy density and pressure, the cosmological energy density and pressure, the dilatonic energy density and pressure. The numerical values in the parameters $\alpha=GM_{\odot}/c^2$ and $\beta=4\pi\epsilon_0/M_{\odot}c^2$ are selected in such a way that the radius $r$ is in kilometers and the mass $m$ is in solar masses.
\par
The boundary conditions in the center of the star are
\begin{equation}
m(0)=0,\quad p(0)=p_c, \quad \Phi(0)=\Phi_c, \quad p_{\Phi}(0)=-\frac{2}{3}\Big( p_c-\frac{\epsilon_c}{3}\Big)-\frac{\Lambda}{3\alpha\beta}V_{,\Phi}(\Phi_c) .
\end{equation}
The boundary conditions on the edge of the star are determined by the condition  $p^{\ast} =p(r^{\ast} ;p_c,\Phi_c)$ (and $\epsilon^{\ast}=0$). Then
\begin{equation}
m^{\ast}=m(r^{\ast};p_c,\Phi_c),\quad \Phi^{\ast}=\Phi(r^{\ast};p_c,\Phi_c),\quad p^{\ast}_{\Phi}=p_{\Phi}(r^{\ast};p_c,\Phi_c).
\label{outbon}
\end{equation}
The radius of a neutron star for physically sound equations of state varies: $r^{\star}\sim 5\div 20$ km.
\par
Outside of the star, where $p\equiv0$ and $\epsilon\equiv0$ we have a dilaton sphere, or a dilasphere. The structure is determined by a shortened system \eqref{MDGM1}$\div$\eqref{MDGPPhi1}. Equation \eqref{MDGP1} is omitted. In the exterior domain we use \eqref{outbon} as left boundary conditions. The right boundary conditions are defined by the cosmological horizon $r_U:$ $\Delta(r_U;p_c,\Phi_c)=0$ where the de Sitter vacuum is reached: $\Phi(r_U;p_c,\Phi_c)=1$.

\section*{2. Results for polytropic equation of state}

The polytropic equation of state we used to solve the equations \eqref{MDGM1}-\eqref{MDGPPhi1} for neutron SSSS in MDG has the following form  $p=Kn_{0}m_{b}(n/n_{0})^{\Gamma}$, $\epsilon=nm_{b}+Kn_{0}m_{b}(n/n_{0})^{\Gamma}/(\Gamma-1)$. We choose  $m_{b}=1.66$ x $10^{-24}$ g and $n_{0}=0.1$ fm$^{-3}$. The value for the parameters $\Gamma$ and $K$ are chosen, so that the solutions are close to the results from the realistic equation of state (Diaz Alonso, 1985). The values $\Gamma =2.34$ and $K=0.0195$ are used to fit equation II from (Diaz Alonso, 1985), as done in (Damour, 1993).The chosen equation of state is suitable, because the maximum mass of a neutron star calculated with it in GR is close to the maximum observed mass of a neutron star (Demorest, 2010), (Antoniadis, 2013)
%\cite{Demorest}, \cite{Antoniadis}.

\begin{figure}[htbp]
\centering
\begin{tabular}{cc}
\subfloat[The mass is in solar masses, the radius in kilometers.]{\epsfig{file=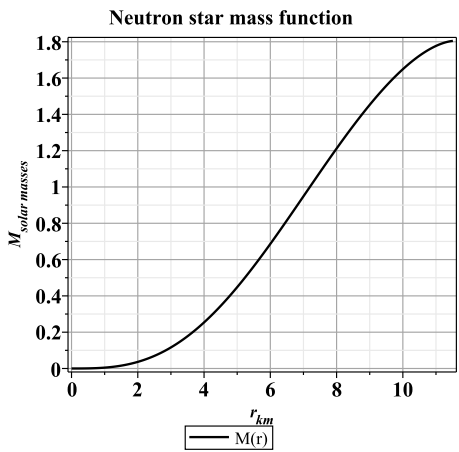, width=0.47\textwidth}} \quad
\subfloat[The pressure is in $10^{34}$ pascals, the radius in kilometers.]{\epsfig{file=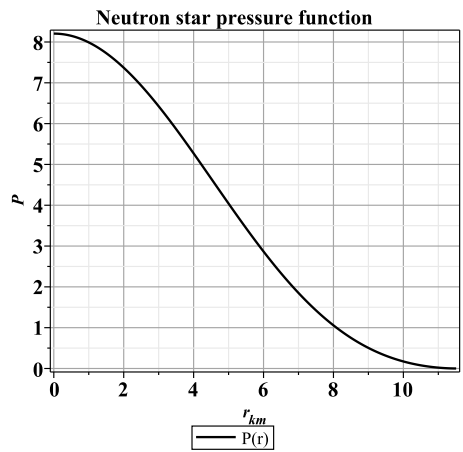, width=0.47\textwidth}}
\end{tabular}
\caption[Mass and pressure.]{On the left is the mass as a function of the radius. On the right is the pressure as function of the radius.}
\label{fig:mass_press}
\end{figure}

\begin{figure}[htbp]
\centering
\begin{tabular}{cc}
\subfloat[The radius in kilometers.]{\epsfig{file=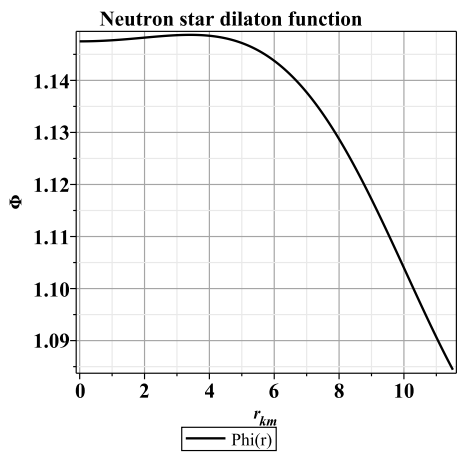, width=0.47\textwidth}}  \quad
\subfloat[The dilaton pressure is in $10^{33}$ pascals, the radius in kilometers.]{\epsfig{file=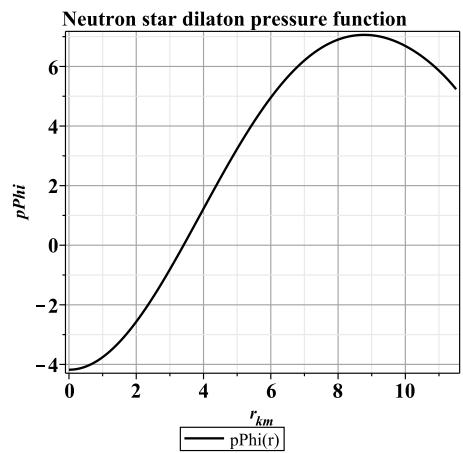, width=0.47\textwidth}}
\end{tabular}
\caption[Dilaton and dilaton pressure.]{On the left is the dilaton as a function of the radius. On the right is the dilaton pressure as a function of the radius.}
\label{fig:dil_dipress}
\end{figure}

\begin{figure}[htbp]
\centering
\begin{tabular}{cc}
\subfloat[$p_{\Lambda }$, $\epsilon_{\Lambda }$ are in unit $10^{31} erg/cm^3$. The radius is in kilometers.]{\epsfig{file=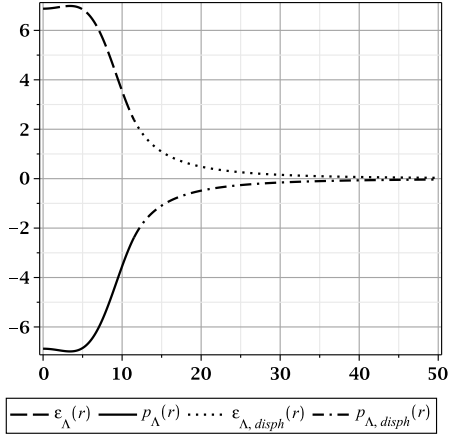, width=0.47\textwidth}} \quad
\subfloat[$ p_{\Phi }$ and $\epsilon_{\Phi }$ are in unit $ 10^{34} erg/cm^3$. The radius is in kilometers.]{\epsfig{file=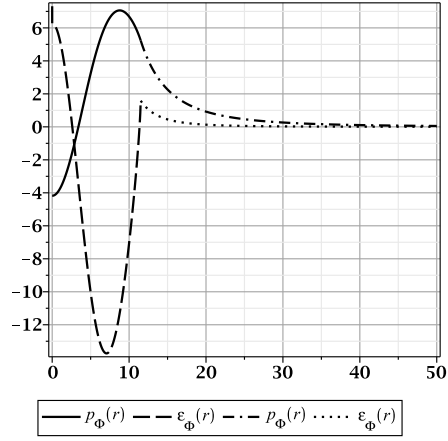, width=0.47\textwidth}}
\end{tabular}
\caption[Energy density and pressure]{On the left are the cosmological energy density and cosmological pressure as functions of the radius for the star and the dilasphere. On the right are the dilaton energy density and cosmological pressure as functions of the radius for the star and the dilasphere.}

\label{fig:p_eps}
\end{figure}

\begin{figure}[htbp]
\centering
\begin{tabular}{cc}
\subfloat[The mass is in solar masses, the radius in kilometers.]{\epsfig{file=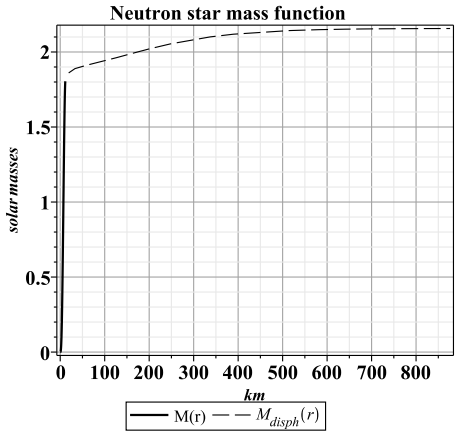, width=0.47\textwidth}} \quad
\subfloat[The pressure is in pascals]{\epsfig{file=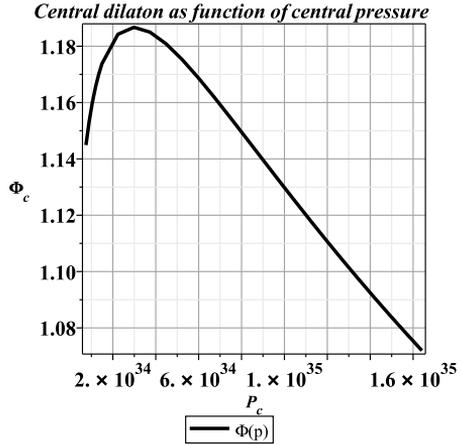, width=0.47\textwidth}}
\end{tabular}
\caption[Mass and pPhi function]{On the left is the mass of the star and the mass of the dilasphere as a function of the radius. On the right is the relation between the central dilaton value and the central pressure.}
\label{fig:mass_phip}
\end{figure}

\begin{figure}[htbp]
	\begin{center}
	\epsfig{file=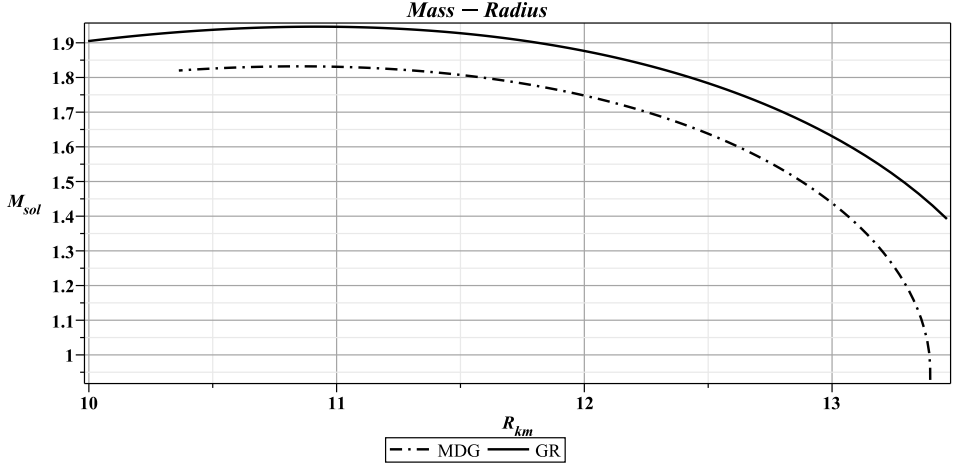, width=0.97\textwidth}
	\caption{The mass-radius relation. The mass is in solar masses, the radius in kilometers. With the lighter color is the result from general relativity. With the darker color is the result from minimal dilatonic gravity for the star without the dilasphere. \label{mass_radius_compare}}
	\end{center}
\end{figure}

\begin{figure}[htbp]
	\begin{center}
	\epsfig{file=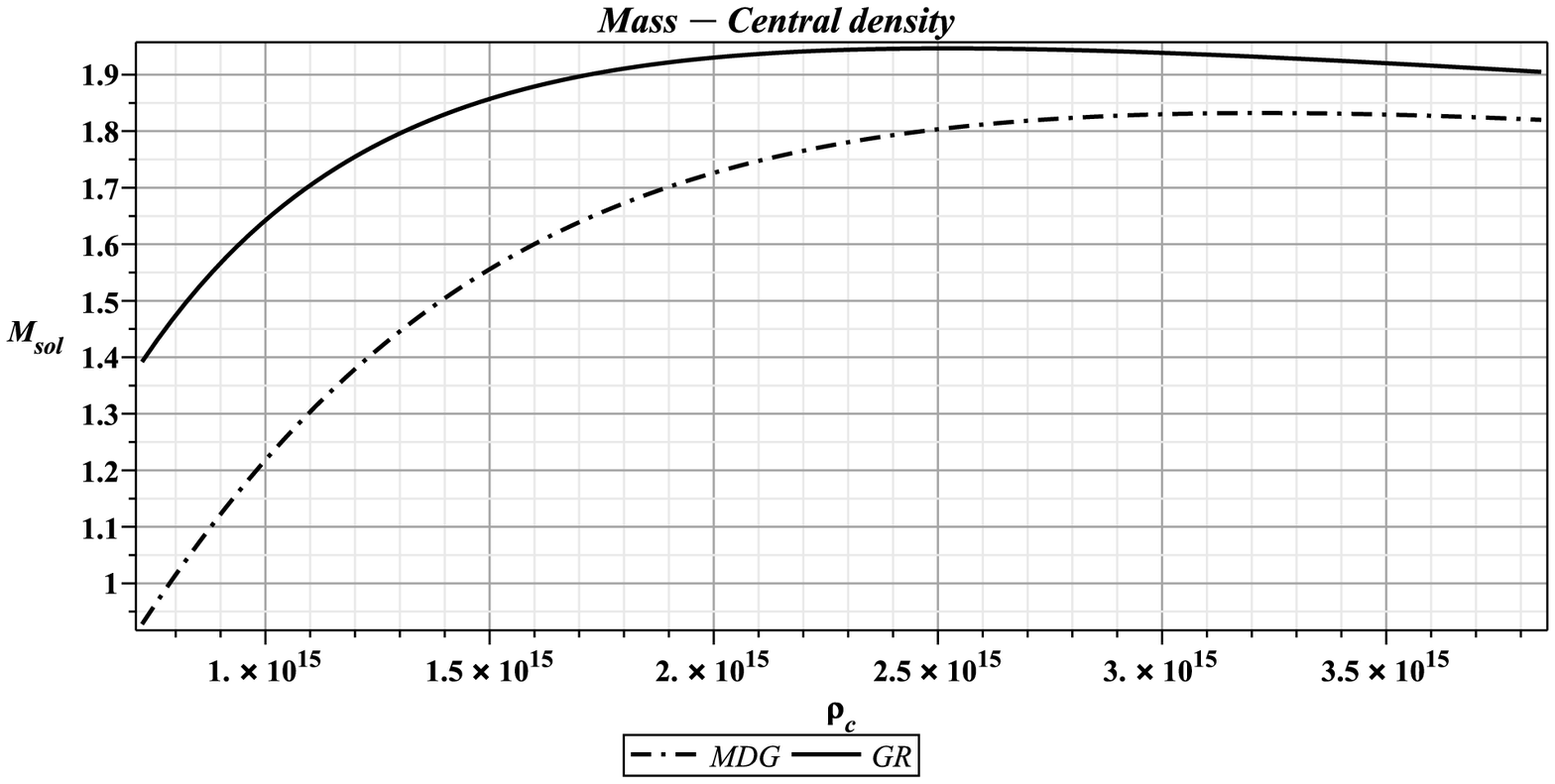, width=0.97\textwidth}
	\caption{The mass-central density relation. The mass is in solar masses, the central density is in $g/cm^3$. With the coral color is the result from general relativity. With the dark orange color is the result from minimal dilatonic gravity for the star without the dilasphere.\label{Mass-central_density}}
	\end{center}
\end{figure}

Figures 1 and 2 describe the internal structure of the neutron star. The results for the mass and pressure are not essentially different from the results in GR. The behaviour of the new, for the model, variable, the dilaton $\Phi$, must be carefully examined on Fig. 2(a). The dilaton is decreasing near the edge of the star, as expected, but in the core of the star there is an increase of the dilaton. Since the quantity 1/Phi describes the variations of the gravitational factor, which describes the changes of the intensity of gravity, we see that it decreases around the center of the star by some 15$\%$. This very interesting result is consistent with the one of articles (Fiziev, 2014a,b). The increase is observed only when the central density is great enough. When the central density is less, the dilaton function decreases from the center of the star to the edge. Why it is happening like that can be answered with the help of Fig. 3(b). When the central density is great enough, the central value of the dilaton pressure $p_{\Phi}$ becomes negative, which leads to increase of the dilaton $\Phi$. When $p_{\Phi}(r)$ becomes positive, then $\Phi(r)$ decreases. Smaller central densities lead to a positive value for $p_{\Phi}$ in the center of the star, and it is a decreasing function from the center to the edge of the star.

\par

A very important feature of the model can be seen on Fig. 4(a). The mass of the dilasphere is clearly seen. In our case under consideration, the mass of the pure neutron star in the MDG model is $M^{\ast}\approx 1.805M_{\odot}$. The dilasphere mass is $M_{disph}\approx 0.352M_{\odot}$ and the mass of the whole object is $M_{total}\approx 2.157M_{\odot}$, i.e. the dilasphere is $\approx 19.5\%$ of the mass of the pure neutron star, or $\approx 16.3\%$ of the mass of the whole object.

This result is also consistent with an analogous one in (Fiziev 2014a,b),
as well as with the results of a more recent article (Capozziello, 2014).
The last paper considers another example of EOS - the simples one for quark stars, and takes into account the mass of the dilasphere.
As a result, the perturbative approach was proved  to be neither inadequate nor incomplete, as wrongly stated in some other papers.

\par

Figure 4(b) shows the important relation between the dilaton in the center of the star and the central pressure. We use that relation to calculate neutron stars with physically different central conditions. Figure 5 shows the mass-radius relation. As we see, the mass of the pure neutron star in MDG model is less than the mass of the star in GR. But the mass of the full object, neutron star and dilasphere in the MDG is greater than the star in GR. We calculated the mass for different central densities (Fig.6). Here we must note that the maximum mass in the MDG model is reached with greater central density.

\par

\section*{3. Conclusion}
In the present article, we study the effects of minimal dilatonic gravity on the structure of static spherically symmetric neutron stars with the polytropic equation of state, consistent with more realistic ones. In general, the obtained results confirm qualitatively the results for similar stars with a more idealized equation of state for ideal Fermi gas at zero temperature studied in (Fiziev, 2014a,b). Some more diverse details in the stars structures were obtained in the case of the polytropic equation of state. The MDG-neutron-stars considered in the present paper, are consistent with the observed neutron stars with total masses about two solar masses. In contrast to the corresponding GR models, an important new feature of the MDG-ones is the existence of a specific dilaton sphere outside the star. It plays the role of dark matter halo and carries about 15 -20$\%$ of the total mass of the object.
We also observe clear signatures of dark energy inside and outside of the star.

The basic conclusion is that the minimal dilatonic gravity model is able to describe realistic models of neutron stars together with the effects of dark matter and dark energy. Its specific feature is that despite the fact that such effects are very small outside the star at scales of stars systems like solar one, these effects become essential inside the star. The physical reason is clear. In physical vacuum outside the star where $\Phi \approx 1$ the cosmological term in the action (1) is extremely small at the scales of stars systems and becomes essential at cosmological scales due to the extremely small value of the cosmological constant $\Lambda$. In contrast, inside the matter $\Phi >1$ and we have huge values of the cosmological potential which compensates the small values of the cosmological constant. This specific novel minimal-dilatonic-gravity-mechanism is simpler than the chameleon one and gives an alternative explanation of the behaviour of dilaton field inside the matter.
\par

\section*{Acknowledgments}

The authors are deeply indebted to Professors Sergei Blinnikov and Andrey Yudin for careful reading of the
manuscript, pointing of an important misprint in the polytropic EOS in the initial text, as well as for useful
suggestions for improvements of the text and for further work. Their idea of checking the existence of
dilasphere using more realistic EOS for a binary pulsar system seems to be promising and deserves special attention
if the dilaton mass has a real value similar to one used in the present preliminary study for purely technical reasons.

Plamen Fiziev is also grateful to Professor Salvatore Capozziello for useful discussion of the basic modern problems of
the modified theories of gravity, especially the ones, related with the star physics and with the article (Capozziello, 2014).

This research was supported in part by the Foundation for Theoretical and Computational
Physics and Astrophysics and Grant of the Bulgarian Nuclear Regulatory Agency for 2014, as
well as by "NewCompStar", COST Action MP1304.

\section*{Authors' contributions}

The present research was performed in close collaboration with two authors.
Fiziev was operative in derivation of the basic equations for
static spherically symmetric star configurations, proposed the topic
and the choice of the polytropic equation. He supervised the performance
of the technical details and their interpretation,
as well as the final version of the text.
Marinov was operative in obtaining all numerical results,
graphs and  in writing the text of the paper.
He also checked all basic formulas and corrected
the important numerical factor 1/3 in the formula for $p_{\Lambda}$ \eqref{9}.

%\clearpage


\begin{thebibliography}{99}
\bibitem[2013]{Antoniadis} Antoniadis \& Co., 2013, {\em Science}, 340 (6131): 1233232
\bibitem[2014]{Capozziello} Astashenok A. V., Capozziello S., Odintsov S. D., 2014, arXiv:1412.5453v1
\bibitem{Clifton} Clifton T., Ferreira P., Padill A., Skordis C., 2012, {\em Physics Reports} 513 1
\bibitem{Damour} Damour T., Esposito-Farense G., 1993, {\em Phys. Rev. Lett.}, 70 15
\bibitem{Defelice} De Felice A., Tsujikawa S., 2010, {\em Living REv. Rel.} 13 3
\bibitem{Demorest} Demorest \& Co., 2010, {\em Nature}, 467 (7319):1081-1083
\bibitem{Diaz} Diaz Alonso J., Ibanez Cabanell J., 1985, {\em ApJ}, 291:308-318
\bibitem{Esp} Esposito-Farese~G., Polarski~D.,2001, {\em Phys. Rev. D} 63 063504.
\bibitem[2006]{Faraoni} Faraoni V., 2006, {\em Phys. Rev. D} 74 104017
\bibitem[2000]{MDG} Fiziev~P., 2000, {\em Mod. Phys. Lett. A} 15, 32, pp. 1977-1990, arXiv:gr-qc/9911037
\bibitem[2000]{mmm} Fiziev P., Yazadjiev S., Boyadjiev T., Todorov M., 2000, {\em Phys. Rev. D}, 61, 124018.
\bibitem[2002]{MDG2} Fiziev P., 2002, arXiv:gr-qc/0202074v4
\bibitem[2003]{4D} Fiziev P., Georgieva D., 2003, {\em Phys. Rev. D} 67 064016.
\bibitem[2013]{MDG1} Fiziev~P., 2013,{\em Phys. Rev. D }87, 044053, arXiv:1209.2695v1 [gr-qc]
\bibitem[2014a]{MDG3} Fiziev P., 2014a, arXiv:1402.2813v1 [gr-qc]
\bibitem[2014b]{MDG3} Fiziev P., 2014b, arXiv:1411.0242v1
\bibitem{Frolov} Frolov A., 2008, {\em PRL} 101 061103
\bibitem{Landau} Landau L., Lifshitz E., 1975, {\em The Classical Theory of Fields}, New York: Pergamon Press
\bibitem[2007]{Nojiri} Nojiri S.,  Odintsov S.D., 2007, Int. J. Geom. Meth. Mod. Phys.4: 115-146, arXiv:hep-th/0601213.
\bibitem[2011]{Nojiri} Nojiri S.,  Odintsov S.D., 2011, Phys. Rept. 505: 59-144, arXiv:1011.0544.
\bibitem[1972]{Ohan} O’Hanlon, 1972, {\em Phys. Rev. Lett}. 29 137
\bibitem{Sotiriou} Sotiriou T., Faraoni V., 2010, {\em REv. Mod. Phys.} 82 451
\bibitem{Starobinsky} Starobinsky A., 1980, {\em Phys.Lett.B} 91 99
\bibitem{Starobinsky1} Starobinsky A., 2007, {\em JEPT Lett.} 86 157

\end{thebibliography}
\end{document}